%% file: TOP-25-016_temp.tex
\documentclass[11pt,twoside,a4paper,cmspaper,final,collab]{cms-tdr}
\def\svnVersion{474da8b}\def\svnDate{2026/01/29}\def\cmsCernNoTag{CERN-EP-2025-287}\def\cmsCernDate{\today}\def\cmsMessage{Published in Physical Review D (Rapid Communications) as \href{http://dx.doi.org/10.1103/hkgw-dztk}{\doi{10.1103/hkgw-dztk}.}}
\begin{document}\cmsNoteHeader{TOP-25-016}

\newcommand{\Mtop}{\ensuremath{m_{\PQt}}\xspace}
\newcommand{\tpt}{\ensuremath{\pt(\PQt)}\xspace}
\newcommand{\ttm}{\ensuremath{m(\ttbar)}\xspace}
\newcommand{\absct}{\ensuremath{\abs{\cos(\theta)}}\xspace}
\newcommand{\ttmcth}{\mbox{\ttm \vs \absct}\xspace}
\newcommand{\tptcth}{\mbox{\tpt \vs \absct}\xspace}
\newcommand{\MINNLO}{\textsc{minnlo}\xspace}
\newcommand{\PWGHRW}{\POWHEG{}+\HERWIG}
\newcommand{\PWGPYT}{\POWHEG{}+\PYTHIA}
\newcommand{\MGHPYT}{\MGvATNLO{}+\PYTHIA}
\newcommand{\MINPYT}{\MINNLO{}+\PYTHIA}
\newcommand{\PWGPYTETA}{\POWHEG{}+\PYTHIA{}+\etat}
\newcommand{\gluglu}{\ensuremath{\Pg\Pg}\xspace}
\newcommand{\srho}{\ensuremath{S(\rho)}\xspace}
\newcommand{\grho}{\ensuremath{\gamma(\rho)}\xspace}
\newcommand{\DE}{\ensuremath{\Delta_{\mathrm{E}}}\xspace}
\newcommand{\etat}{{\HepParticle{\PGh}{\PQt}{}}\xspace}

\newcommand{\cpe}{\ensuremath{C_{\perp}}\xspace}
\newcommand{\cpa}{\ensuremath{C_{\parallel}}\xspace}

\newcommand{\phim}{\ensuremath{\ket{\Phi^-}}\xspace}
\newcommand{\psip}{\ensuremath{\ket{\Psi^+}}\xspace}
\newcommand{\psim}{\ensuremath{\ket{\Psi^-}}\xspace}
\newcommand{\phipm}{\ensuremath{\ket{\Phi^\pm}}\xspace}
\newcommand{\psipm}{\ensuremath{\ket{\Psi^\pm}}\xspace}
\newcommand{\ket}[1]{\ensuremath{\big| {#1} \big\rangle}\xspace}
\newcommand{\bra}[1]{\ensuremath{\big\langle {#1} \big|}\xspace}

\cmsNoteHeader{TOP-25-016}
\title{Characterization of the quantum state of top quark pairs produced in proton-proton collisions at \texorpdfstring{$\sqrt{s}=13\TeV$}{sqrt(s)=13 TeV} using the beam and helicity bases}

\abstract{Measurements of the spin correlation coefficients in the beam basis are presented for top quark-antiquark (\ttbar) systems produced in proton-proton collisions at $\sqrt{s}=13\TeV$ collected by the CMS experiment in 2016--2018, and corresponding to an integrated luminosity of 138\fbinv. The \ttbar system is reconstructed from final states containing an electron or muon, and jets. Together with the previously reported results in the helicity basis, these measurements are used to decompose the system into the Bell and spin eigenstates in various kinematic regions. The spin correlation coefficients are also used to evaluate properties of the \ttbar quantum state, such as the purity, von Neumann entropy, and entanglement. All results are consistent with standard model predictions.}

\date{\today}

\hypersetup{
pdfauthor={CMS Collaboration},
pdftitle={Characterization of the quantum state of top quark pairs produced in proton-proton collisions at sqrt(s)=13 TeV using the beam and helicity bases},
pdfsubject={CMS},
pdfkeywords={CMS, top quark pairs, quantum properties}}

\maketitle

The top quark is the most massive known fundamental particle, with a lifetime on the order of $10^{-25}\unit{s}$. This lifetime is shorter than both the quantum chromodynamics (QCD) hadronization time scale $1/\Lambda_{\text{QCD}} \approx 10^{-24}\unit{s}$ and the spin decorrelation time scale $\Mtop/\Lambda_{\text{QCD}}^2 \approx 10^{-21}\unit{s}$, where \Mtop denotes the top quark mass~\cite{ParticleDataGroup:2024, Mahlon:2010gw}. Consequently, the top quark spin information is preserved in the angular distribution of its decay products, which makes top quark-antiquark (\ttbar) pairs excellent probes for studying polarization and spin correlations~\cite{Bernreuther:2008ju, Bernreuther:2015yna}. The spins of the top quarks are expected to be strongly correlated~\cite{Baumgart:2012twist}, which can translate into the \ttbar pair having nonclassical correlations, \ie, entangled top quarks.

The recent measurements of polarization and spin correlations in the helicity basis, as well as the observation of entanglement, by the ATLAS~\cite{ATLAS:2023fsd} and CMS~\cite{CMS-PAS-TOP-23-001, CMS-TOP-23-007} Collaborations, mark a significant step in the characterization of the \ttbar system. In this letter, we extend the measurements of the spin correlation matrix to the beam basis. While the helicity basis is particularly suited for analyzing the spin correlations and entanglement at high invariant masses of the \ttbar system, \ttm, the beam basis provides complementary sensitivity to the measurements near the production threshold~\cite{Bernreuther:2004jv, Cheng:2024btk}.

The CMS apparatus~\cite{CMS:2008xjf,CMS:2023gfb} is a multipurpose, nearly hermetic detector, designed to trigger on~\cite{CMS:2020cmk,CMS:2016ngn,CMS:2024aqx} and identify electrons, muons, photons, and hadrons~\cite{CMSelectrons,CMS:2018rym,CMS:2014pgm}. A global algorithm~\cite{PF} aims to reconstruct all individual particles in an event, combining information provided by the all-silicon tracker and by the crystal electromagnetic and brass-scintillator hadron calorimeters, operating inside a 3.8\unit{T} superconducting solenoid, with data from the gas-ionization muon detectors interleaved with the layers of the flux-return yoke outside the solenoid. The reconstructed particles are used to build \PGt leptons, jets, and missing transverse momentum~\cite{CMS:2018jrd,JET,CMS:2019ctu}.

Collision data taken from 2016 to 2018 and corresponding to 138\fbinv~\cite{CMS:LUM-17-003, CMS:LUM-17-004, CMS:LUM-18-002} have been used for the measurements reported in Ref.~\cite{CMS-TOP-23-007}. Following the conventions introduced in Ref.~\cite{Bernreuther:2015yna} these measurements were used to derive the results in the helicity basis presented in this article. It should be noted that this definition of the helicity basis includes a discrete rotation of $\pi$ around the $k$ axis if $\theta>\pi/2$, where $k$ and $\theta$ are the top quark's direction vector and its scattering angle, respectively, in the \ttbar rest frame. This convention was introduced to account for the Bose--Einstein symmetry of the gluon-gluon (\gluglu) initial state.

In the beam basis, the $z$ axis is aligned with the proton beams, with fixed $x$ and $y$ axes given by the CMS coordinate system, \ie, the $x$ axis pointing toward the center of the LHC ring, and the $y$ axis oriented vertically upward~\cite{CMS:2008xjf}. Similar to the definition of the helicity basis, the top quarks and their decay products are actively boosted from the laboratory frame into the \ttbar rest frame, followed by an active boost of the top (anti)quark decay products from the \ttbar rest frame into their parent's rest frame, where active boosts, used in the definition of both bases, imply Lorentz transformations of the particles' four-momenta while the coordinate systems remain unchanged. As a consequence of the symmetries around the beam axis and between the incoming protons, the spin correlation matrix $C_{ij}$ is diagonal with $\cpe = C_{xx} = C_{yy}$ and $\cpa = C_{zz}$, and the polarization coefficients $P_i$ and $\overline{P}_i$ of the top quark and antiquark, respectively, are equal to zero~\cite{Afik:2020topentangled}. For the beam basis, the additional rotation to account for the Bose--Einstein symmetry is not applied. Since the axes in the helicity basis are defined event-wise based on the top quark direction, it is necessary to reanalyze each selected event to obtain the results in the beam basis.

At the CERN LHC, the \ttbar system is produced predominantly via \gluglu fusion. Quark-antiquark annihilation also contributes, accounting for approximately 10\% of the production rate on average. Near the production threshold, the $\gluglu\to\ttbar$ events are produced predominantly in a spin-singlet configuration. In contrast, at high transverse momentum of the top quark, \tpt, the production approaches a spin-triplet state~\cite{Afik:2020topentangled}. The contributions of individual states can be evaluated from the density matrix $\rho$, which can be written in the Fano--Bloch decomposition~\cite{Fano:1983zz} as
\begin{equation}\label{eq_densitymatrix}
\ifthenelse{\boolean{cms@external}}{\begin{aligned}
    \rho = \frac{1}{4} \Big( I \otimes I & + \sum_i P_i \sigma_i \otimes I  \\ & + \sum_j \overline{P}_j I \otimes \sigma_j - \sum_{ij}  C_{ij} \sigma_i \otimes \sigma_j \Big),
\end{aligned}}{
    \rho = \frac{1}{4} \Big( I \otimes I + \sum_i P_i \sigma_i \otimes I + \sum_j \overline{P}_j I \otimes \sigma_j - \sum_{ij}  C_{ij} \sigma_i \otimes \sigma_j \Big),
}
\end{equation}
where $I$ is the 2x2 identity matrix and $\sigma_i$ are Pauli matrices.

To characterize the quantum state, we perform the eigenvector decomposition of $\rho$ and identify the eigenvectors in the helicity basis with the Bell states~\cite{Nielsen:2012yss}
\begin{equation}\begin{aligned}
    \phipm &= \frac{1}{\sqrt{2}} \left( \ket{\uparrow\uparrow} \pm \ket{\downarrow\downarrow} \right) \\
    \psipm &= \frac{1}{\sqrt{2}} \left( \ket{\uparrow\downarrow} \pm \ket{\downarrow\uparrow} \right),
\end{aligned}\end{equation}
and in the beam basis with the spin states: singlet \psim (pseudoscalar) and triplet \ket{\uparrow\uparrow}, \psip,
\ket{\downarrow\downarrow} (vector).

For the spin quantization axis we choose the $k$ direction in the helicity basis and the $z$ direction in the beam basis. The contributions of the individual states are evaluated as the expectation value of $\rho$ for the corresponding state \ket{\psi_i}
\begin{equation}
    \bra{\psi_i} \rho \ket{\psi_i}.
\end{equation}
In quantum mechanics, the purity is defined as
\begin{equation}
    \grho = \Tr\big({\rho}^2\big),
    \label{eq_purity}
\end{equation}
which measures how mixed a quantum state is. For a pure state, $\rho^2 = \rho$, and thus purity equals unity. In contrast, for mixed states $\Tr\big(\rho^2\big) < 1$.

The von Neumann entropy \srho, which is a quantum mechanical generalization of the Shannon entropy~\cite{Fano:1957zz}, characterizes the amount of uncertainty about the value of a random variable, and is defined as
\begin{equation}
    \srho = -\Tr\big(\rho \log_2{\rho}\big).
    \label{eq_vonNeumann}
\end{equation}
The entropy is zero only for pure states and reaches its highest value for the maximally mixed states, following the intuition of the classical entropy. Since the \ttbar system involves two spin-$1/2$ particles, the entropy range is $0 < \srho < 2$. Purity and entropy quantify whether the \ttbar system is described predominantly by a single state or by a mixture of several states. High purity (or low entropy) indicates that a single state predominantly contributes, whereas low purity (or high entropy) signals that several states contribute comparably.

A sufficient condition for the entanglement based on the Peres--Horodecki criterion~\cite{Peres:1996sep, Horodecki:1997sep} is given by the entanglement marker~\cite{Afik:2020topentangled,Afik:2022kwm}
\begin{equation}
    \DE=C_{11}+\abs{C_{22}+C_{33}}>1,
    \label{method_eq3}
\end{equation}
where the indices $(1,2,3)$ of the diagonal elements of the spin correlation matrix refer to $(x,y,z)$ in the beam and $(n,r,k)$ in the helicity basis. This condition applies under the assumption that the \ttbar state is fully described by quantum mechanics~\cite{Abel:1992kz, Han:2023fci}.

{\tolerance=1200
We present the characterization of the quantum properties of the \ttbar system in bins of \ttmcth and \tptcth. In the helicity basis, the polarizations and spin correlations are directly taken from Ref.~\cite{CMS-TOP-23-007}, while in the beam basis, the measurements of \cpe and \cpa are new results. This analysis uses the same method introduced in Ref.~\cite{CMS-TOP-23-007}, including the object selection, reconstruction of the \ttbar system, and treatment of systematic uncertainties. The polarization and spin correlation coefficients are determined from a template fit to the angular distributions of the \ttbar decay products. Since only \cpe and \cpa can be nonzero in the beam basis, all other coefficients are fixed to zero in the fit.
\par}

The results are compared to the values predicted by the standard model using the Monte Carlo simulated samples described in Ref.~\cite{CMS-TOP-23-007}. These predictions are obtained using the matrix element (ME) event generator \POWHEG \textsc{v2}~\cite{Nason:2004rx,Frixione:2007vw,Frixione:2007nw} subsequently combined with the parton shower simulation from \PYTHIA8.240 \cite{Sjostrand:2014zea}, using the underlying event tune CP5~\cite{CP5}. For these predictions we evaluate the ME scale and parton distribution function (PDF) uncertainties. In addition, the measured coefficients in the beam basis are compared to the predictions obtained using \PWGHRW7.1~\cite{Bahr:2008pv} with tune CH3~\cite{CMS:2020dqt}, and \MINPYT, a \ttbar sample at next-to-next-to-leading order (NNLO) QCD generated with \POWHEG \MINNLO~\cite{Mazzitelli_2021} in combination with the CP5 \PYTHIA tune. Tabulated results for all figures, including the \tptcth measurements, are provided in the HEPData record for this analysis~\cite{hepdata}.

\begin{figure*}[hp!]
    \centering
    \includegraphics[width=0.48\textwidth]{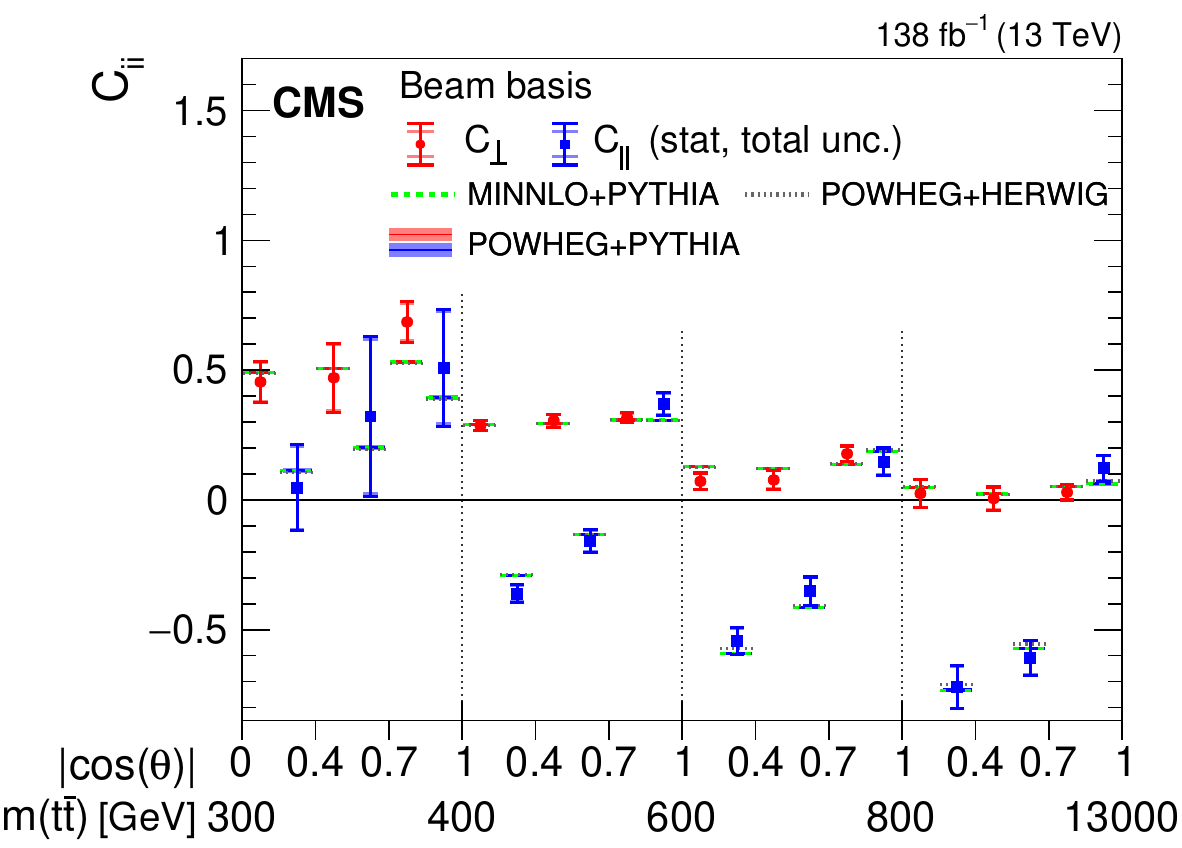}
    \hfill
    \includegraphics[width=0.48\textwidth]{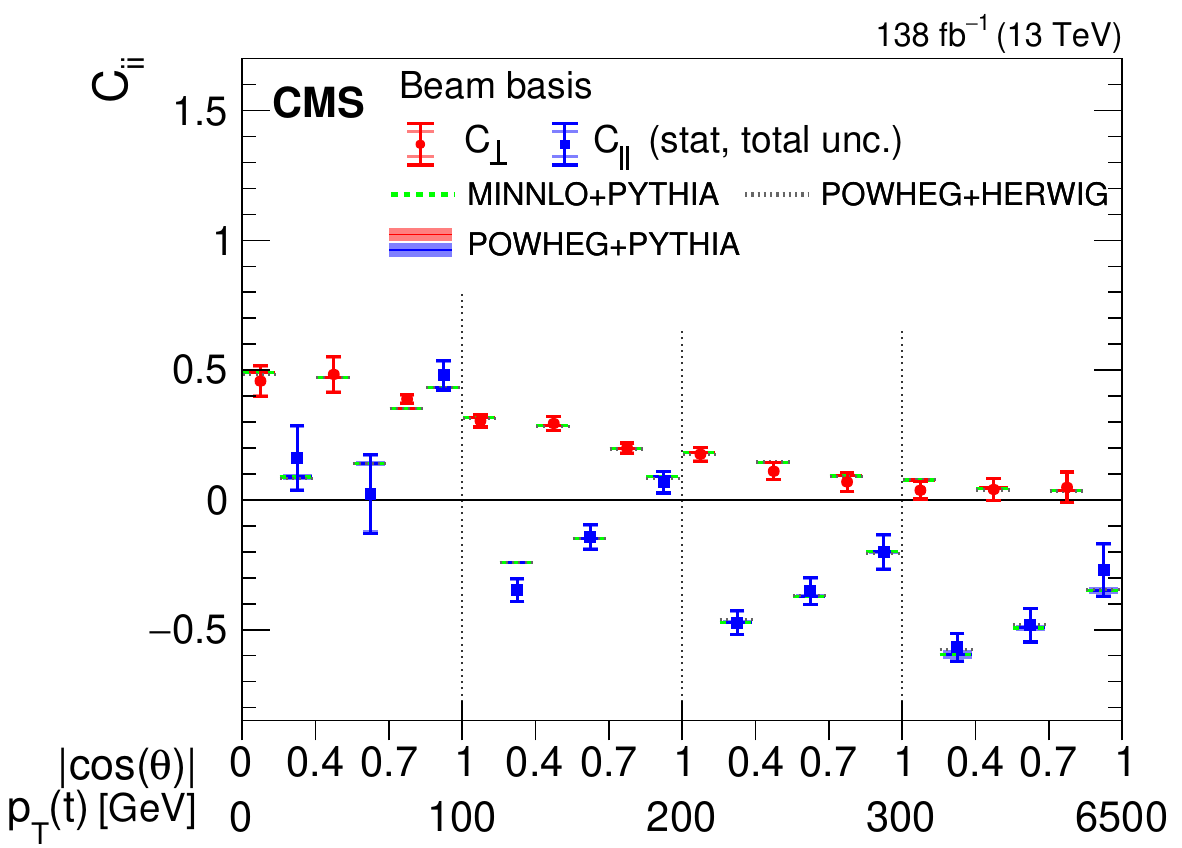}
    \caption{Results of the spin correlation coefficients in bins of \ttmcth (left) and \tptcth (right) in the beam basis. The measurements (markers) are shown with the statistical (inner error bars) and total (outer error bars) uncertainties and compared to the predictions from \PWGHRW, \MINPYT, and \PWGPYT. The \PWGPYT prediction is displayed with the ME scale and PDF uncertainties.}
    \label{results_coeffs_beambasis}
\end{figure*}

{\tolerance=20000
In Fig.~\ref{results_coeffs_beambasis} we present the measured spin correlation coefficients in the beam basis in \ttmcth and \tptcth bins. The statistical uncertainties shown as the inner error bars dominate and nearly coincide with the total uncertainties shown as the outer error bars. From $\chi^2$ tests comparing measured coefficients to the various predictions, the $p$-values range between 0.22--0.34 (0.70--0.79) for the \ttmcth (\tptcth) measurements, indicating agreement in all cases. These $p$-values are obtained taking into account the full experimental covariance matrix. The corresponding coefficients for the helicity basis can be found in Ref.~\cite{CMS-TOP-23-007}.
\par}

The uncertainties in the observables derived from the polarization and spin correlation coefficients are evaluated using pseudo-experiments, which are generated by sampling from a multivariate normal distribution defined by the measured central values and the covariance matrix of the coefficients. To ensure physically meaningful results, pseudo-experiments are rejected if they lead to negative eigenvalues of the density matrix or if any of the spin correlation coefficients satisfy $\abs{C_{ij}} > 1$. Potential biases in the pseudo-experiments have been evaluated and are negligible relative to the measurement uncertainties. The statistical uncertainties dominate and nearly coincide with the total uncertainties in all the observables.

\ifthenelse{\boolean{cms@external}}{\begin{figure*}[p!]}{\begin{figure*}[t!]}
    \centering
    \includegraphics[width=0.48\textwidth]{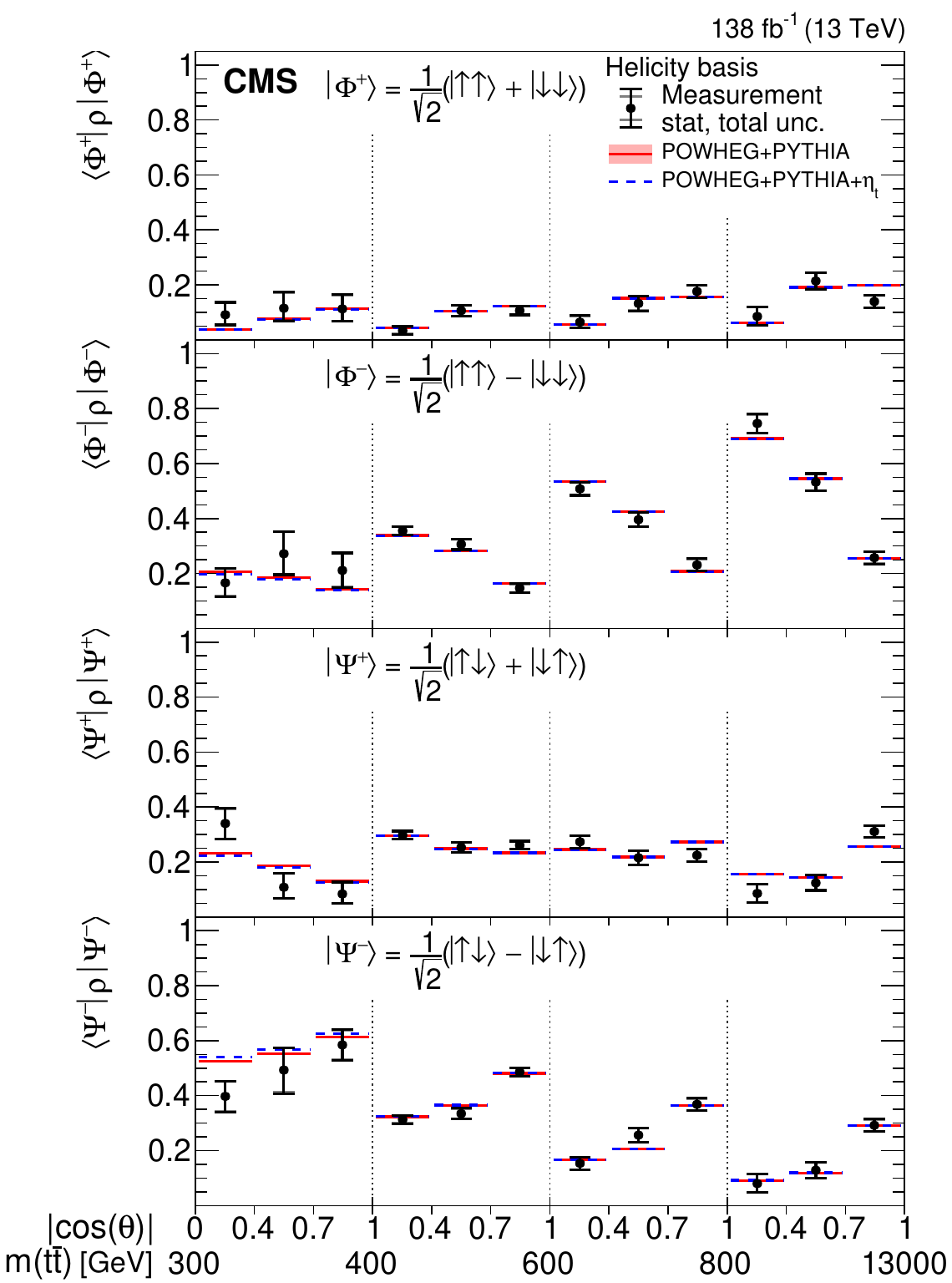}
    \hfill
    \includegraphics[width=0.48\textwidth]{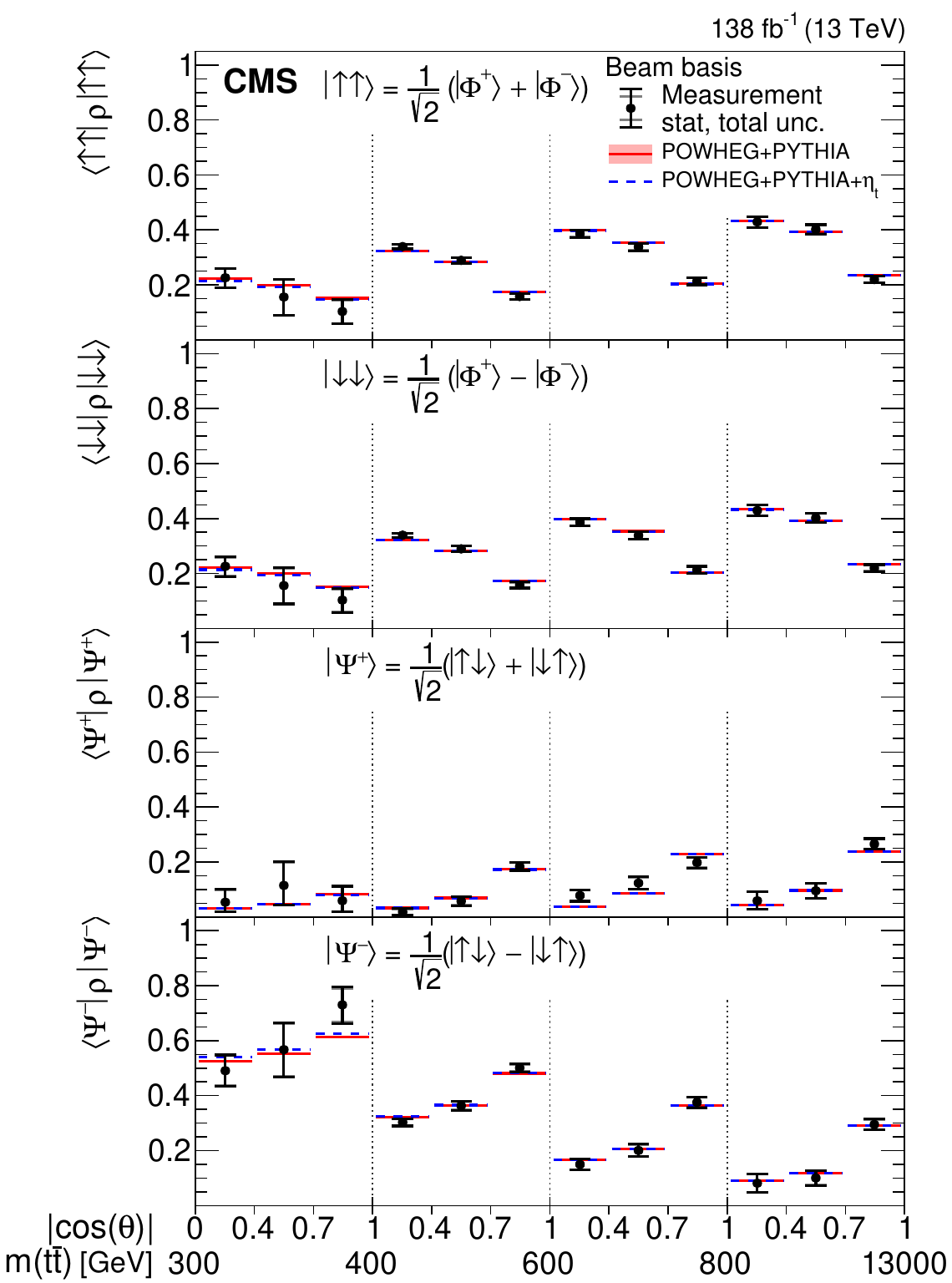}
    \caption{Results of the state decomposition in terms of the Bell and spin states in bins of \ttmcth in the helicity (left) and beam (right) basis. The measurements (markers) are shown with the statistical (inner error bars) and total (outer error bars) uncertainty and compared to the predictions from \PWGPYT, and \PWGPYTETA. The \PWGPYT prediction is displayed with the ME scale and PDF uncertainties.}
    \label{resultsdata_decompo}
\end{figure*}

The eigenvector decomposition of $\rho$ is shown in Fig.~\ref{resultsdata_decompo} and compared to the \PWGPYT simulation. In the helicity basis for $\ttm>800\GeV$ and $\absct<0.4$, we see the largest individual contribution ($\sim$70\%) of the \phim Bell state. In both the helicity and the beam basis, high contributions (60--70\%) of the \psim pseudoscalar state are observed near the \ttbar production threshold for $\ttm<400\GeV$. Although the short lifetime of the top quark prevents the formation of long-lived \ttbar bound states, nonrelativistic QCD calculations predict bound-state-like enhancements near the \ttbar production threshold~\cite{Fadin1990, Kiyo2009, SuminoYokoya2010, Ju:2020otc, Fuks:2021xje}. These effects are simulated using \MGHPYT as a pseudoscalar particle, \etat, produced via \gluglu fusion with a mass of 343\GeV and a width equal to twice the top quark width, which decays as $\etat \to \PWp \PQb \PWm \PAQb$ with the same spin correlations as \ttbar pairs in a pure \psim spin-singlet state. This simulation is normalized to the cross section of 6.43\unit{pb} from Ref.~\cite{Fuks:2021xje} and added to the nominal \PWGPYT prediction as an additional comparison in Fig.~\ref{resultsdata_decompo}. The expected enhancement of the \psim state caused by the \etat contribution near the threshold is smaller than the uncertainty, hence the two models cannot be distinguished by these measurements. The contributions of the triplet states \ket{\uparrow\uparrow} and \ket{\downarrow\downarrow} are equal in the beam basis because of the symmetry between the incoming protons.

\ifthenelse{\boolean{cms@external}}{\begin{figure*}[p!]}{\begin{figure*}[t!]}
    \centering
    \includegraphics[width=0.48\textwidth]{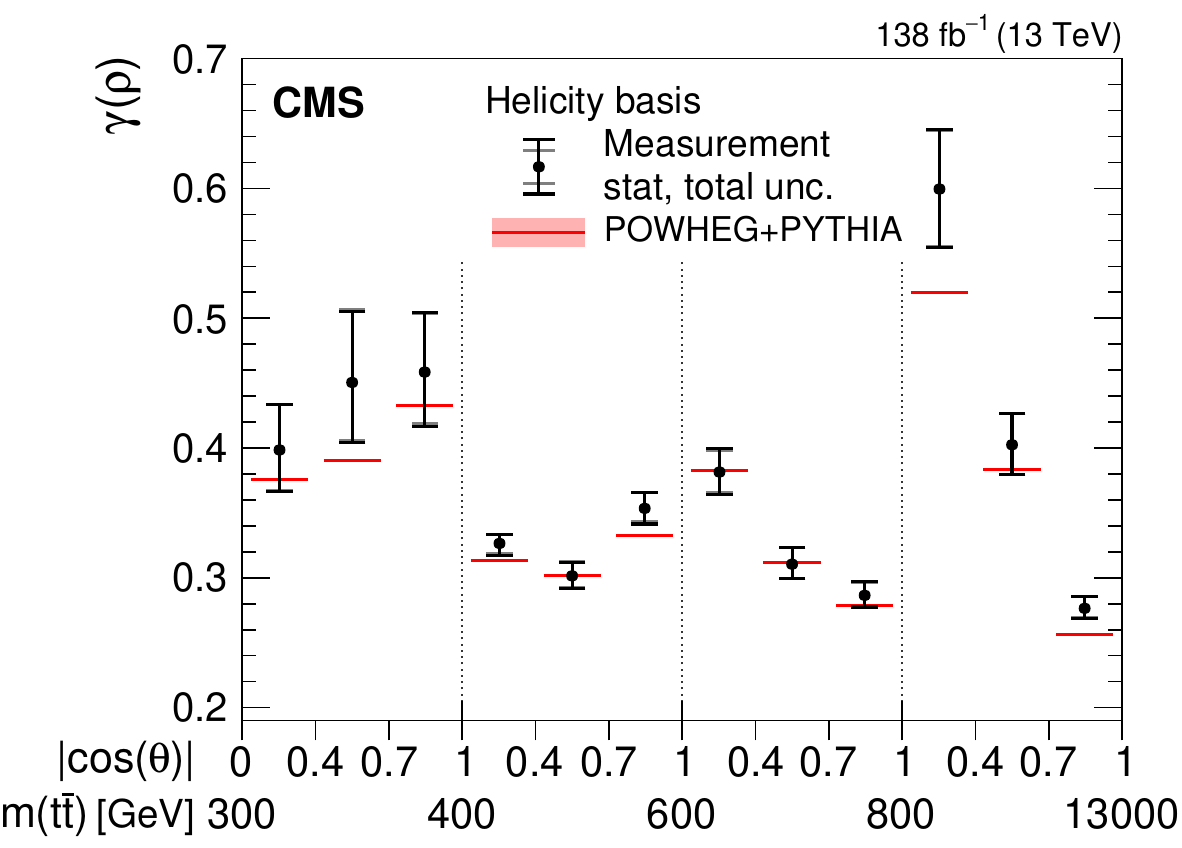}
    \hfill
    \includegraphics[width=0.48\textwidth]{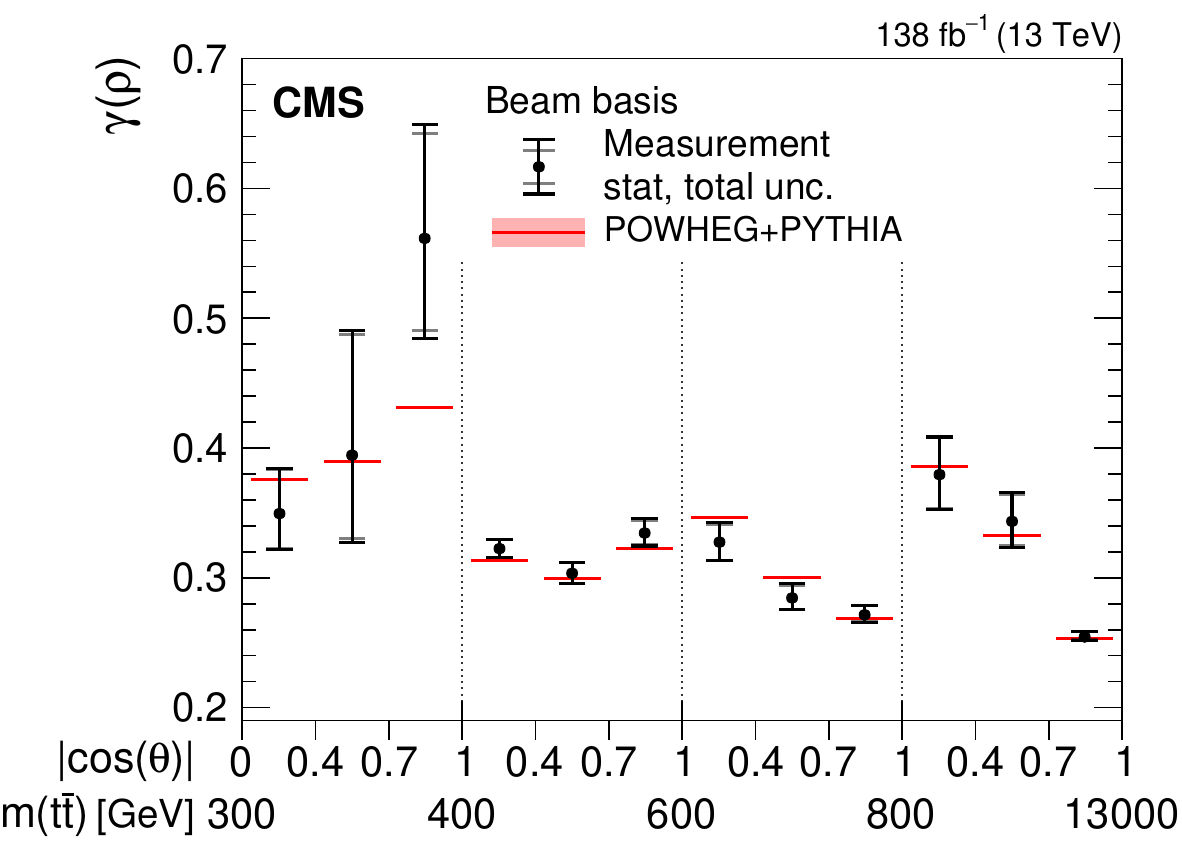} \\[1ex]
    \includegraphics[width=0.48\textwidth]{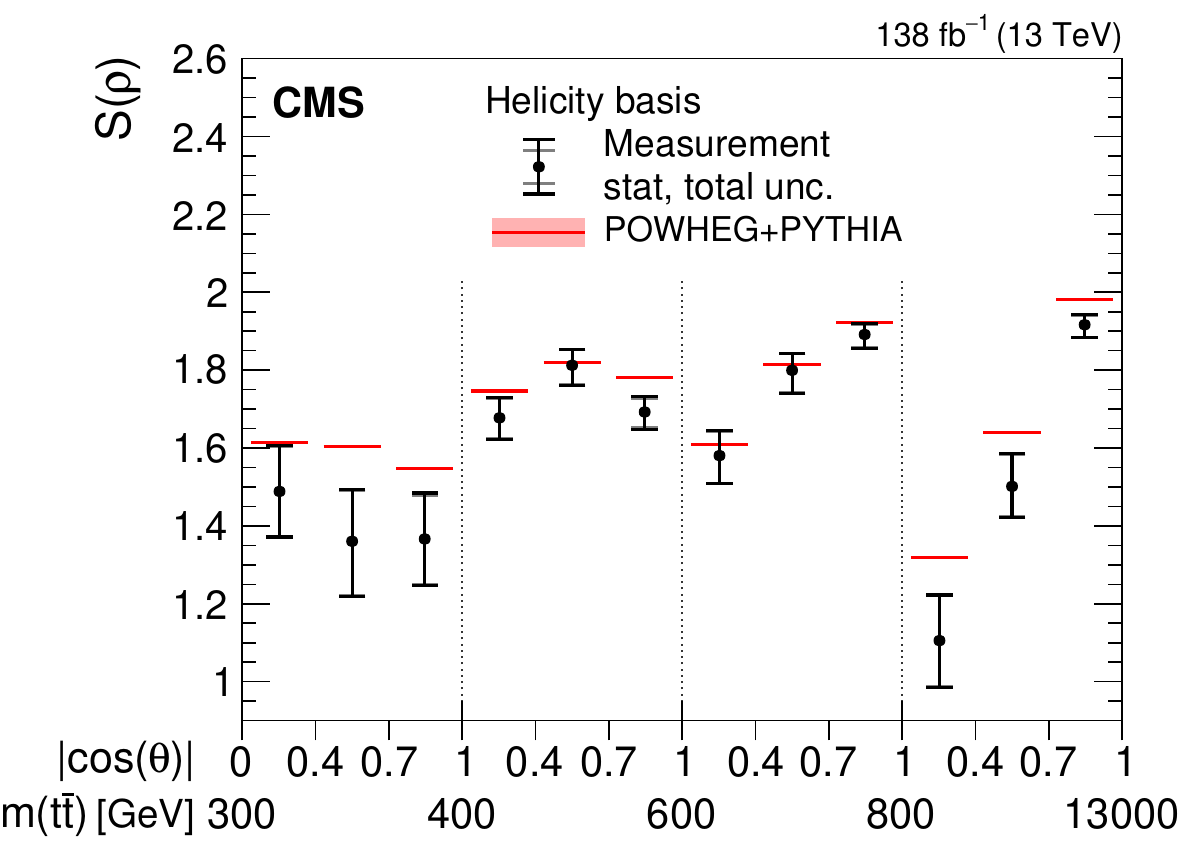}
    \hfill
    \includegraphics[width=0.48\textwidth]{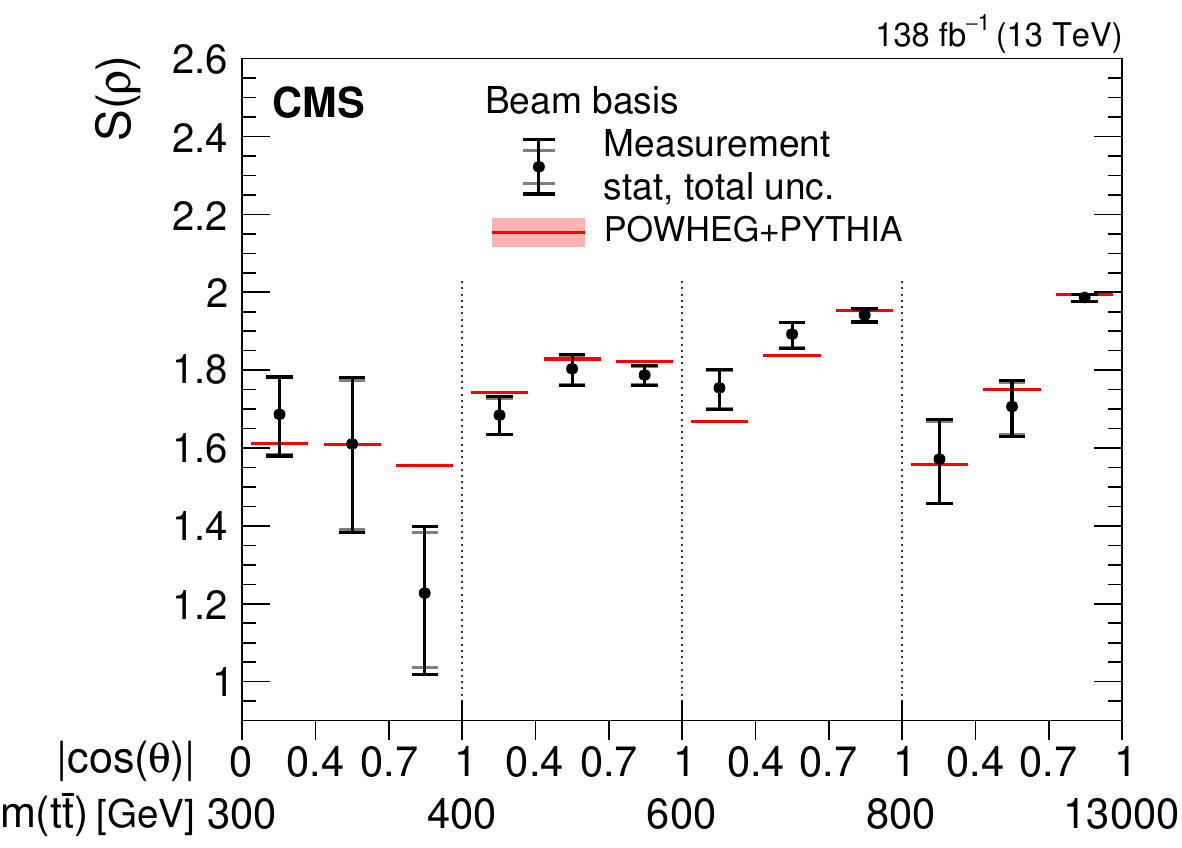}
    \caption{Results of the purity \grho (upper) and entropy \srho (lower) measurements in bins of \ttmcth in the helicity (left) and beam (right) basis. The measurements (markers) are shown with the statistical (inner error bars) and total (outer error bars) uncertainty and compared to the predictions from \PWGPYT, shown with the ME scale and PDF uncertainties.}
    \label{resultsdata_purityentropy}
\end{figure*}

In Fig.~\ref{resultsdata_purityentropy} the extracted purity \grho and entropy \srho values are shown in the helicity and beam bases. Overall, we observe agreement with the standard model predictions for both \grho and \srho. The maximum deviations are found in the helicity basis for the lowest and highest \ttm bins, where a slightly higher \grho and lower \srho than expected are observed. However, the differences are all within 2 standard deviations. In the helicity basis, \grho is highest at high \ttm and low \absct, where \srho reaches its minimum value. In the beam basis, \grho is highest at low \ttm and high \absct, which corresponds to the region of lowest \srho. As expected, \srho is nonzero across all bins. We note that in Fig.~\ref{resultsdata_decompo} for both bases in the $\ttm>800\GeV$, $\absct>0.7$ bin, all the Bell or spin states make similar contributions, which is reflected in the purity being the lowest in this bin.

In Fig.~\ref{resultsdata_entang}, the entanglement marker \DE values are shown in the helicity and beam bases. The dashed green line represents the entanglement criterion: states with $\DE > 1$ are entangled. In the helicity basis, the \DE is highest at high \ttm and low \absct, where the significance of entanglement exceeds 5 standard deviations, as was already reported in Ref.~\cite{CMS-TOP-23-007}. In the beam basis we find evidence of entanglement in the threshold region, with a significance of greater than 3 standard deviations for $\absct > 0.7$.

\begin{figure*}[!tp]
    \centering
    \includegraphics[width=0.48\textwidth]{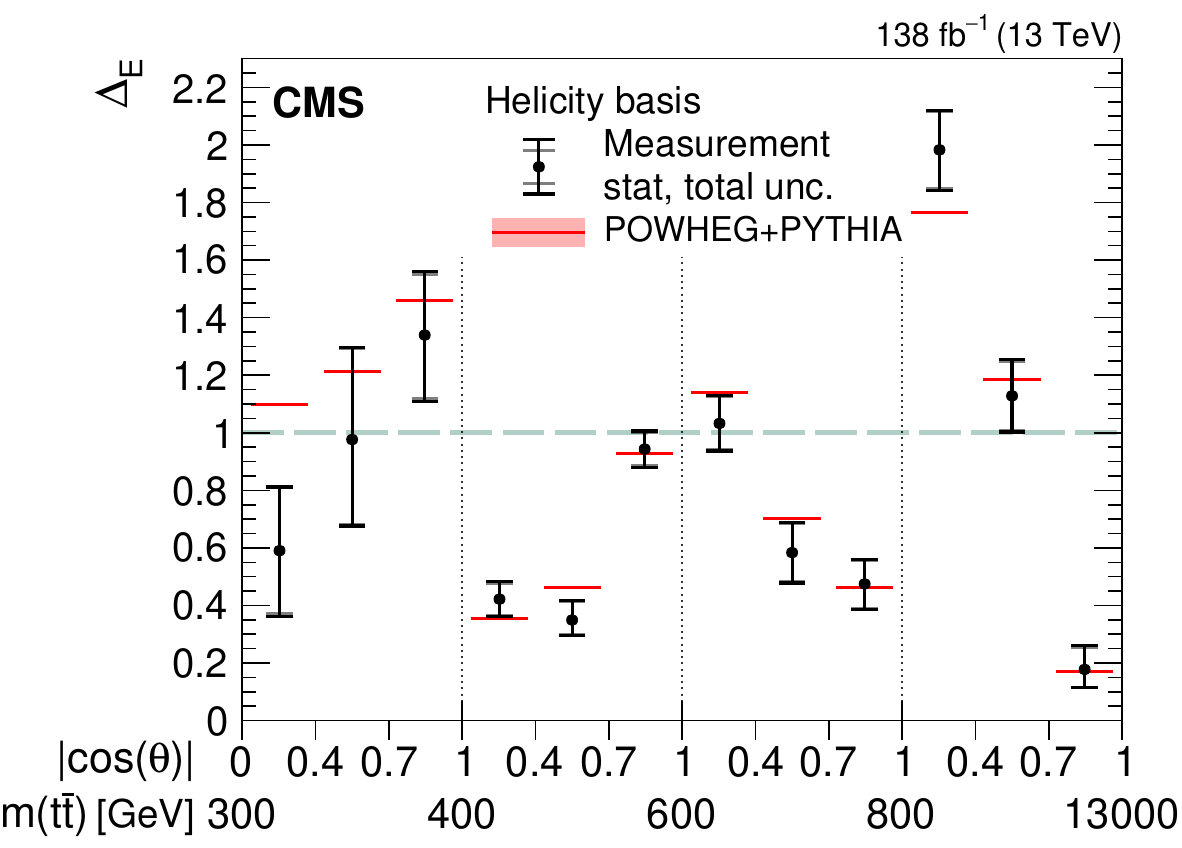}
    \hfill
    \includegraphics[width=0.48\textwidth]{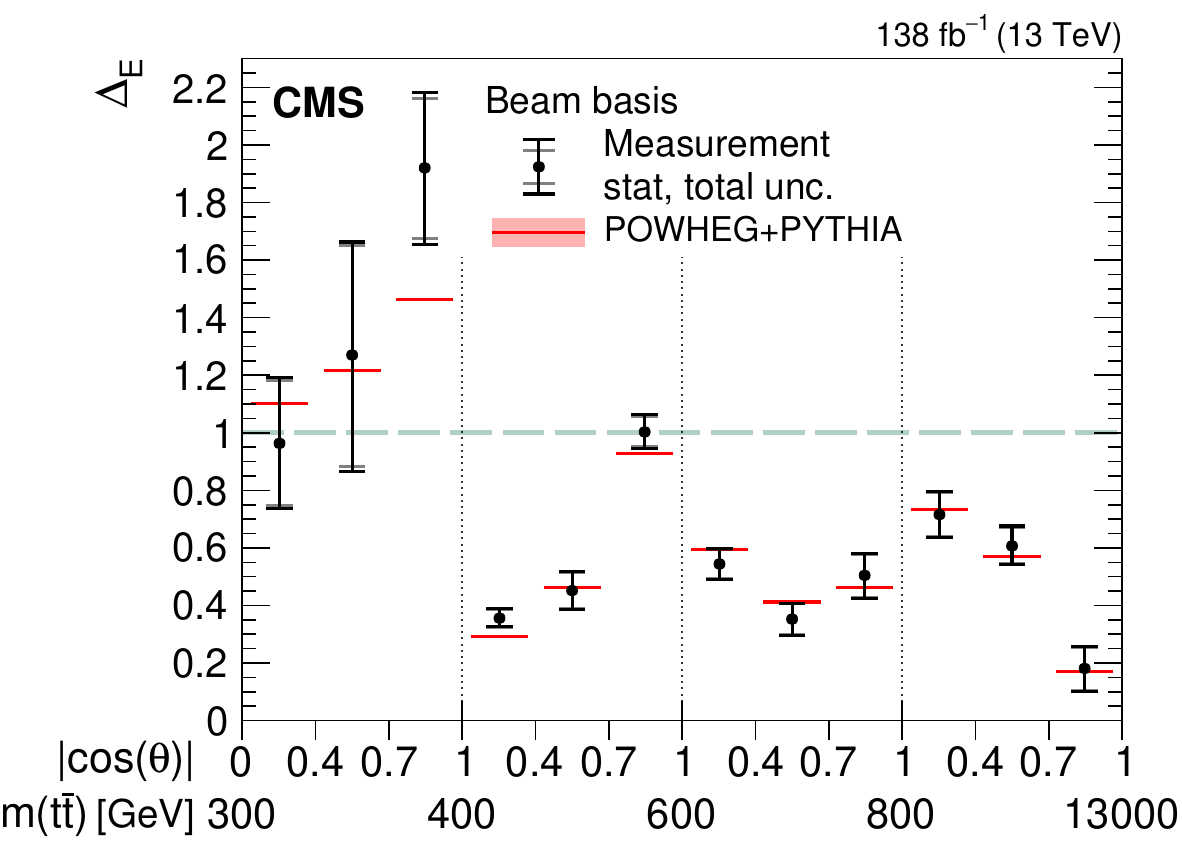}
    \caption{Results of the entanglement marker \DE measurements in bins of \ttmcth in the helicity (left) and beam (right) basis. The measurements (markers) are shown with the statistical (inner error bars) and total (outer error bars) uncertainty and compared to the predictions from \PWGPYT, shown with the ME scale and PDF uncertainties. The dashed green line represents the lower bound for entangled states.}
    \label{resultsdata_entang}
\end{figure*}

In summary, the spin correlation measurements of the top quark-antiquark (\ttbar) system from Ref.~\cite{CMS-TOP-23-007} are extended to include results in the beam basis for the first time. This is done with bins in two dimensions, comprising the top quark scattering angle and either the invariant mass of the \ttbar system or the transverse momentum of the top quark, using proton-proton collisions at $\sqrt{s}=13\TeV$ recorded by the CMS experiment at the LHC, corresponding to an integrated luminosity of 138\fbinv. The density matrix is decomposed into eigenstates, which are identified as Bell states in the helicity basis and spin states in the beam basis. We also present the first experimental results for the purity, von Neumann entropy, and entanglement of the \ttbar system in both bases. All the measurements are consistent with standard model expectations.

\begin{acknowledgments}
We congratulate our colleagues in the CERN accelerator departments for the excellent performance of the LHC and thank the technical and administrative staffs at CERN and at other CMS institutes for their contributions to the success of the CMS effort. In addition, we gratefully acknowledge the computing centers and personnel of the Worldwide LHC Computing Grid and other centers for delivering so effectively the computing infrastructure essential to our analyses. Finally, we acknowledge the enduring support for the construction and operation of the LHC, the CMS detector, and the supporting computing infrastructure provided by the following funding agencies: SC (Armenia), BMBWF and FWF (Austria); FNRS and FWO (Belgium); CNPq, CAPES, FAPERJ, FAPERGS, and FAPESP (Brazil); MES and BNSF (Bulgaria); CERN; CAS, MoST, and NSFC (China); MINCIENCIAS (Colombia); MSES and CSF (Croatia); RIF (Cyprus); SENESCYT (Ecuador); ERC PRG, TARISTU24-TK10 and MoER TK202 (Estonia); Academy of Finland, MEC, and HIP (Finland); CEA and CNRS/IN2P3 (France); SRNSF (Georgia); BMFTR, DFG, and HGF (Germany); GSRI (Greece); NKFIH (Hungary); DAE and DST (India); IPM (Iran); SFI (Ireland); INFN (Italy); MSIT and NRF (Republic of Korea); MES (Latvia); LMTLT (Lithuania); MOE and UM (Malaysia); BUAP, CINVESTAV, CONACYT, LNS, SEP, and UASLP-FAI (Mexico); MOS (Montenegro); MBIE (New Zealand); PAEC (Pakistan); MES, NSC, and NAWA (Poland); FCT (Portugal); MESTD (Serbia); MICIU/AEI and PCTI (Spain); MOSTR (Sri Lanka); Swiss Funding Agencies (Switzerland); MST (Taipei); MHESI (Thailand); TUBITAK and TENMAK (T\"{u}rkiye); NASU (Ukraine); STFC (United Kingdom); DOE and NSF (USA).

\hyphenation{Rachada-pisek} Individuals have received support from the Marie-Curie programme and the European Research Council and Horizon 2020 Grant, contract Nos.\ 675440, 724704, 752730, 758316, 765710, 824093, 101115353, 101002207, 101001205, and COST Action CA16108 (European Union); the Leventis Foundation; the Alfred P.\ Sloan Foundation; the Alexander von Humboldt Foundation; the Science Committee, project no. 22rl-037 (Armenia); the Fonds pour la Formation \`a la Recherche dans l'Industrie et dans l'Agriculture (FRIA) and Fonds voor Wetenschappelijk Onderzoek contract No. 1228724N (Belgium); the Beijing Municipal Science \& Technology Commission, No. Z191100007219010, the Fundamental Research Funds for the Central Universities, the Ministry of Science and Technology of China under Grant No. 2023YFA1605804, the Natural Science Foundation of China under Grant No. 12061141002, 12535004, and USTC Research Funds of the Double First-Class Initiative No.\ YD2030002017 (China); the Ministry of Education, Youth and Sports (MEYS) of the Czech Republic; the Shota Rustaveli National Science Foundation, grant FR-22-985 (Georgia); the Deutsche Forschungsgemeinschaft (DFG), among others, under Germany's Excellence Strategy -- EXC 2121 ``Quantum Universe" -- 390833306, and under project number 400140256 - GRK2497; the Hellenic Foundation for Research and Innovation (HFRI), Project Number 2288 (Greece); the Hungarian Academy of Sciences, the New National Excellence Program - \'UNKP, the NKFIH research grants K 131991, K 133046, K 138136, K 143460, K 143477, K 146913, K 146914, K 147048, 2020-2.2.1-ED-2021-00181, TKP2021-NKTA-64, and 2025-1.1.5-NEMZ\_KI-2025-00004 (Hungary); the Council of Science and Industrial Research, India; ICSC -- National Research Center for High Performance Computing, Big Data and Quantum Computing, FAIR -- Future Artificial Intelligence Research, and CUP I53D23001070006 (Mission 4 Component 1), funded by the NextGenerationEU program (Italy); the Latvian Council of Science; the Ministry of Education and Science, project no. 2022/WK/14, and the National Science Center, contracts Opus 2021/41/B/ST2/01369, 2021/43/B/ST2/01552, 2023/49/B/ST2/03273, and the NAWA contract BPN/PPO/2021/1/00011 (Poland); the Funda\c{c}\~ao para a Ci\^encia e a Tecnologia, grant CEECIND/01334/2018 (Portugal); the National Priorities Research Program by Qatar National Research Fund; MICIU/AEI/10.13039/501100011033, ERDF/EU, "European Union NextGenerationEU/PRTR", and Programa Severo Ochoa del Principado de Asturias (Spain); the Chulalongkorn Academic into Its 2nd Century Project Advancement Project, the National Science, Research and Innovation Fund program IND\_FF\_68\_369\_2300\_097, and the Program Management Unit for Human Resources \& Institutional Development, Research and Innovation, grant B39G680009 (Thailand); the Kavli Foundation; the Nvidia Corporation; the SuperMicro Corporation; the Welch Foundation, contract C-1845; and the Weston Havens Foundation (USA).
\end{acknowledgments}
\bibliography{auto_generated}
\cleardoublepage \appendix\section{The CMS Collaboration \label{app:collab}}\begin{sloppypar}\hyphenpenalty=5000\widowpenalty=500\clubpenalty=5000\input{TOP-25-016-public-authorlist.tex}\end{sloppypar}
\end{document}

%% file: TOP-25-016-public-authorlist.tex
\cmsinstitute{Yerevan Physics Institute, Yerevan, Armenia}
{\tolerance=6000
A.~Hayrapetyan, V.~Makarenko\cmsorcid{0000-0002-8406-8605}, A.~Tumasyan\cmsAuthorMark{1}\cmsorcid{0009-0000-0684-6742}
\par}
\cmsinstitute{Institut f\"{u}r Hochenergiephysik, Vienna, Austria}
{\tolerance=6000
W.~Adam\cmsorcid{0000-0001-9099-4341}, L.~Benato\cmsorcid{0000-0001-5135-7489}, T.~Bergauer\cmsorcid{0000-0002-5786-0293}, M.~Dragicevic\cmsorcid{0000-0003-1967-6783}, P.S.~Hussain\cmsorcid{0000-0002-4825-5278}, M.~Jeitler\cmsAuthorMark{2}\cmsorcid{0000-0002-5141-9560}, N.~Krammer\cmsorcid{0000-0002-0548-0985}, A.~Li\cmsorcid{0000-0002-4547-116X}, D.~Liko\cmsorcid{0000-0002-3380-473X}, M.~Matthewman, J.~Schieck\cmsAuthorMark{2}\cmsorcid{0000-0002-1058-8093}, R.~Sch\"{o}fbeck\cmsAuthorMark{2}\cmsorcid{0000-0002-2332-8784}, M.~Shooshtari\cmsorcid{0009-0004-8882-4887}, M.~Sonawane\cmsorcid{0000-0003-0510-7010}, W.~Waltenberger\cmsorcid{0000-0002-6215-7228}, C.-E.~Wulz\cmsAuthorMark{2}\cmsorcid{0000-0001-9226-5812}
% ...
\par}
\vskip\cmsinstskip
\dag:~Deceased\\
$^{1}$Also at Yerevan State University, Yerevan, Armenia\\
$^{2}$Also at TU Wien, Vienna, Austria\\
% ...
